# A Sense of Crisis: Physics in the *fin-de-siècle* Era[*]

HELGE KRAGH
Centre for Science Studies, Department of Physics and Astronomy
Aarhus University, 8000 Aarhus, Denmark
E-mail: helge.kragh@ivs.au.dk

**Abstract** Whereas physics in the period from about 1880 to 1910 experienced a steady growth, it was also a revolutionary period in which the foundations of the physical world picture were criticized and reconsidered. Generally speaking, from about 1890 mechanics and materialism came under increasing attack and sought replaced by new formulations based on either energy, the ether, or the electromagnetic field. *Fin-de-siècle* physics was in many ways a chapter of turmoil in the history of science. I review the main developments and alternatives to the established physics, in particular energetics, ether physics, the electromagnetic world view, and also the role played by radioactivity and other new rays discovered in the years around 1900. In the end the anticipated revolution based on the "matter is dead" catchword did not succeed. A revolution did take place in the period, but it was a different one that did not derive from the Zeitgeist of *fin de siècle*.

Physics at the turn of the nineteenth century was a highly developed, professional science. But it was a small science, worldwide comprising between 1,200 and 1,500 academic physicists, almost all of them from Europe or North America (Kragh 1999: 13-22). Although international it was not global. Experimental and applied aspects dominated the field, yet the last decades of the century witnessed the emergence and growth of theoretical physics as distinct from the earlier tradition of mathematical physics (Jungnickel and McCormmach 1986).

---

[*] Contribution to Michael Saler, ed., *The Fin-de-Siècle World*, to be published by Routledge in 2014.

The *fin-de-siècle* period from about 1890 to 1905 saw several attempts at establishing a new, modern foundation of physics, but what today is known as modern physics – essentially relativity and quantum physics – had other roots. The notion of a "classical" physics and the classical-modern dichotomy first appears in 1899, referring to mechanics (Staley 2005). Only a decade later did physicists begin to talk about the modern versus the classical physics in the sense used today. When used in the *fin-de-siècle* period, the term "modern" physics typically referred to either programs of the future or to alternatives to the mechanical foundation of physics such as the electron theories of the early twentieth century.

**The mechanical world view**

It is often assumed that the revolution in the physical sciences that occurred in the early twentieth century was preceded by a period in which the Victorian generation of physicists complacently accepted the supremacy of mechanics in all of science. Physics, so they are to have believed, was and would remain solidly based on Newton's mechanical laws and the forces acting between bodies, either long-range forces as in astronomy and electricity or short-range forces as in atomic and molecular physics. According to the philosopher and mathematician Alfred North Whitehead, the last quarter of the nineteenth century was a period of what Thomas Kuhn many years later would call normal science bound by a paradigm. It was "an age of successful scientific orthodoxy, undisturbed by much thought beyond the conventions," even "one of the dullest stages of thought since the time of the First crusade" (Whitehead 1925: 148). However, Whitehead's characterization is a caricature of *fin-de-siècle* physics, a chapter in the history of science that was anything but dull.

Newtonian mechanics, or rather the more advanced versions of mechanics with roots in Newton's laws, was undoubtedly held in great esteem in the late nineteenth century. It was widely accepted that the goal of physics – or sometimes even its definition – was the reduction of all physical phenomena to the principles of mechanics. There were physicists who believed that their science was essentially complete and that future physics would remain mechanical in its foundation, but this was hardly the generally held view (Badash 1972; Schaffer 2000). Nor was the orthodoxy complete, for the basis in mechanics was not accepted dogmatically or always thought to be universally valid. On the contrary, there was in the 1890s a lively and many-faceted discussion of what the mechanical clockwork universe was, more precisely, and how valid – and desirable – it was. The two new sciences of thermodynamics and electrodynamics, both going back to the mid-century, differed in many ways from classical mechanics, and yet it was widely believed that they could be understood on a mechanical basis, indeed that such an understanding had already been achieved. James Clerk Maxwell saw no contradiction between his field theory of electromagnetism and Newton's theory of mechanics.

On the other hand, even physicists of a conservative inclination recognized that there were a few dark clouds on the mechanical heaven. On 27 April 1900 Lord Kelvin (William Thomson) gave a famous address to the Royal Institution in London on "Nineteenth Century Clouds over the Dynamical Theory of Heat and Light" in which he singled out two problems. One of them was the problem of the stationary ether and the lack of any physical effect of the earth's motion through it. The other and even darker cloud related to the successful kinetic theory of gases as founded on a mechanical basis by Maxwell and the Austrian physicist Ludwig Boltzmann. Briefly, this theory seemed to be consistent only if it were assumed that the molecules or atoms of a gas were rigid bodies with no internal parts. This assumption contradicted the evidence from spectroscopy which strongly indicated that the atom had an internal structure, and a most complex structure at that (Brush 1986: 353-363).

There were other problems, some of them relating to the second law of thermodynamics according to which the entropy of a physical system, a measure of its degree of molecular disorder, increases spontaneously and irreversibly in time. The fundamental entropy law thus expresses a direction of time, from the past to the future, but how can this law possibly be explained on the basis of the time-symmetric laws of mechanics? Although the problem received a solution of sorts with Boltzmann's probabilistic theory of entropy of 1877, it continued to be controversial. Some physicists considered the second law an insurmountable obstacle for the mechanization of nature. For example, in 1896 Ernst Zermelo in Germany concluded that the law flatly disagreed with Newtonian mechanics, which he saw as a problem for the mechanical world view rather than the second law of thermodynamics. According to him, he had demonstrated that matter cannot be composed of discrete particles governed by the laws of mechanics. Boltzmann denied the validity of Zermelo's argument and remained convinced that there was no deep disagreement between mechanics and thermodynamics. He described himself as a classical physicist and his conviction as characteristic of classical physics.

Even Newton's divine law of gravitation, the central element in celestial dynamics and a paragon for the mechanical world view, was not beyond criticism. It had been known since the late 1850s that it could not account with sufficient precision for the motion of Mercury, an anomaly that could only be removed by the addition of arbitrary hypotheses of an ad hoc nature. There were in the period around 1900 several attempts to improve Newton's law or to provide it with a non-mechanical foundation in terms of either hydrodynamics or electrodynamics. Contributors to the latter line of approach, based on either Maxwellian or non-Maxwellian versions of electrodynamics, included Hendrik A. Lorentz in the Netherlands and Wilhelm Scheibner and Richard Gans in Germany. Although these theories were unsuccessful, they indicate a characteristic willingness of *fin-de-siècle* physicists to challenge even the most sacred parts of the mechanical research program. More was soon to follow.

Although the supremacy of mechanics was increasingly questioned, in the 1890s only a minority of physicists saw the situation as a serious crisis in the mechanical world view. They typically responded to the problems and critiques by reformulating mechanics and avoiding from presenting it as an overarching world view. Rather than *explaining* nature mechanically, they retreated to the more modest position of *describing* it mechanically (Heilbron 1982; Seth 2007). Such an attitude was consonant with the popular "descriptionist" idea that science was not concerned with truths, and certainly not with absolute truths. Scientific theories were considered to be just condensed and economic descriptions of natural phenomena. Similarly, there was nothing more to understanding than equations and models.

**A gospel of energy**

While the second law of thermodynamics might cause problems for the mechanical world picture, the consensus view was that energy itself and its associated law of energy conservation (the first law of thermodynamics) were successfully explained in mechanical terms. However, according to some physicists and chemists energy was more fundamental than matter, and thermodynamics more fundamental than mechanics. The physicist Georg Helm and the physical chemist Wilhelm Ostwald, a future Nobel laureate, arrived at this conclusion at about 1890, coining the name *energetics* for their new research program of a unified and generalized thermodynamics (Hiebert 1971; Görs, Psarros, and Ziche 2005). They and their allies promoted energetics as more than just a new scientific theory: it was meant to be an alternative to the existing understanding of nature, which they claimed was a "scientific materialism" based on mechanics and the hypothesis of matter as composed by atoms and molecules. As they saw it, mechanics was to be subsumed under the more general laws of energetics in the sense that Newton's mechanical laws were held to be reducible to energy principles.

Moreover, according to Ostwald and a few other energeticists, the generalized concept of energy was not restricted to physical phenomena but included also mental phenomena such as willpower and happiness (Hakfoort 1992). Generally, elements of *Lebensphilosophie* were parts of Ostwald's version of energetics. While some critics saw this as a problematic sign of materialism, Ostwald insisted that energetics was thoroughly anti-materialistic, a much needed revolt against the dominance of matter in science. He further thought that an energy-based science would eventually result in a better world, both materially and spiritually. This element of utopianism was not restricted to the energetics movement, but can be found also in other sciences at the turn of the century (Kemperink and Vermeer 2010).

For a decade or so the energetics movement occupied an important position in German scientific and cultural life. Although primarily German, its influence extended to other countries, including France, Sweden, Italy and the United States, in many cases as part of the more general positivistic philosophy known as monism. The French physicist and philosopher Pierre Duhem arrived, largely independently, to a vision of a generalized thermodynamics as a non-mechanical, phenomenological theory of everything. His ideas had much in common with those of Ostwald and Helm, yet there were also differences.

For example, whereas the energetics movement was anti-religious or at least anti-clerical, Duhem presented his version of energetics as in harmony with orthodox Catholicism, maintaining that science could not dispense with the notion of matter. He had no taste for Ostwald's extension of thermodynamics beyond the limit of traditional science.

Ostwald and Duhem had in common that they wanted to rid science of visualizable hypotheses and analogies with mechanics, in particular the seductive illusion of atoms and molecules as real material entities. Thermodynamics, they argued, had the great advantage that it was neutral as to the constitution of matter, indeed neutral as to the existence of matter. In this respect energetics agreed with the ideas of the Austrian physicist-philosopher Ernst Mach, who from a positivist perspective held that atoms were nothing but convenient fictions. On the other hand, many scientists considered the controversial anti-atomism of the energetics alternative as reason enough to dismiss it or ignore it. Ostwald's colleagues in chemistry objected that they could not solve their problems without the concept of atoms. While Ostwald eventually conceded that atoms existed, Duhem and Mach went in their graves (both in 1916) without accepting the reality of atoms.

At the annual meeting of the German Association of Natural Scientists and Physicians in Lübeck in 1895, Ostwald gave a programmatic address in which he argued that energetics was destined to be the scientific world view of the future and already on its way to overcome the inherent limitations of scientific materialism. "The most promising scientific gift that the closing century can offer the rising century," he said, "is the replacement of the materialistic world view by the energeticist world view" (Ostwald 1904: 231). In the debate that followed the views of Ostwald and Helm were attacked by Boltzmann, in particular, who denied that he was a scientific materialist in Ostwald's sense of the term. He concluded that the energetics program was without scientific merit and that it was closer to ideology than science. As far as he was concerned, mechanics was still the only science sufficiently developed to secure scientific progress. Max Planck, who a few years later would initiate quantum theory, came to agree with Boltzmann. He considered energetics an unsound and unproductive version of natural philosophy, what he called a dilettantish speculation. Although other German physicists expressed some sympathy with the ideas of energetics, its impact on physics was limited and short-lived. While it made sense to deny the reality of atoms in 1895, ten years later it was a position that was much harder to defend.

Few British physicists had sympathy for the abstract, anti-metaphysical and positivistic view of science championed in different ways by Ostwald, Duhem and Mach. Nor had they any sympathy for the anti-religious tendencies with which energetics and positivism were associated in Germany. In a critical comment of 1896 on Ostwald's energetics program, the Irish physicist George Francis FitzGerald distinguished between the metaphysically receptive British style and the unphilosophical German style that presented science as a "well arranged catalogue of facts without any hypotheses." No, "A Briton wants emotion – something to raise enthusiasm, something with a human interest," and this was sorely absent from the dry view of science – "the culmination of the pessimism of Schopenhauer" – advocated by Ostwald and his allies (Brush 1978: 96).

**The heat death**

Thermodynamics entered *fin-de-siècle* physics also in a cosmological context, primarily in the form of the so-called heat death and its consequences for the past and future of the universe. Shortly after the formulation of the second law of thermodynamics several physicists, among them William Thomson in England and Hermann von Helmholtz in Germany, pointed out that on the assumption that the law was of unrestricted validity it would lead to an irreversible "degradation" of energy throughout the cosmos. The amount of energy would remain constant, but in the far future it would be unable to generate further physical activity of any kind. The stars would no longer shine, and all life would become extinct. In 1867 the German physicist Rudolf Clausius, another of the pioneers of thermodynamics, formulated the pessimistic scenario of a *Wärmetod* in terms of the global entropy tending towards a maximum: the closer the entropy approached this maximum, characterized by a complete disorganization of material structures, the closer would the universe approach an equilibrium state of unchanging death. As Clausius emphasized, the law of entropy increase contradicted the popular view of a cyclic universe in which the conditions constantly recur, yet in such a way that the overall state of the universe remains the same.

Disseminating quickly from the world of theoretical physics to the general cultural arena, the heat death scenario was hotly debated from about 1870 to 1910. Not only was the scientifically based prediction of an end to the world (meaning an end of all activity in the world) highly controversial, it also seemed to follow from it that the world must have had a beginning in time. The "entropic creation" argument first stated about 1870 relies on the simple observation that we live in a low-entropy world. Had the world existed in an eternity of time, entropy would by now have increased to its maximum value, and consequently the age of the world must be finite. There must have been a beginning. To the mind of most people, whether scientists or nonscientists, a cosmic beginning implied creation, and creation implied a creator. In other words, the law of entropy increase could be used apologetically, as a scientific proof of God.

The heat death and the associated notion of cosmic creation was highly controversial and endlessly debated in the last quarter of the nineteenth century (Kragh 2008). The subject was an integral part of the more general cultural and social struggle between, on the one hand, materialists, positivists and socialists, and, on the other, protagonists of the established world order and its belief in religious and spiritual values. As many saw it, reversibility and the mechanical world view was associated with materialism and implicitly atheism while these views were contradicted by the irreversibility expressed by the second law of thermodynamics. In Germany the issue became part of the *Kulturkampf* that raged in the 1870s and 1880s, with many Catholic thinkers arguing for the inevitability of the heat death and, in some cases, using the authority of thermodynamics as an argument for divine creation. Those opposed to such reasoning could easily avoid the conclusions, for example by inventing counter-entropic processes or by claiming that the second law was inapplicable to an infinitely large universe. Or, less reasonably, they could simply deny the validity of the second law, such as did the famous German zoologist and self-fashioned physicist Ernst Haeckel in his enormously popular *Die Welträtsel* from 1899. A prominent advocate of scientism and atheism, Haeckel was a leader of the monist movement to which also Ostwald belonged. He claimed that the universe was infinite and immortal, an uncreated *perpetuum mobile*, and that the law of entropy increase was consequently refuted.

It should be pointed out that the entropic controversy in the late nineteenth century mostly involved philosophers, social critics, theologians and amateur scientists. Although a few physicists and astronomers contributed to the discussion, most stayed silent, convinced that it was of a metaphysical rather than scientific nature. The consensus view among astronomers was that the universe was probably infinite, but they realized that there was no way to prove it observationally and preferred to limit their science to what could be observed by their telescopes. The universe at large did not belong to astronomy, its state of entropy even less so. Generally, the large majority of professional physicists denied that their science was relevant to the larger themes of *fin-de-siècle* ideology such as degeneration and decadence. Occasionally inorganic decay or energy degradation as expressed in the law of entropy increase was associated with the general feeling of degeneration, but such ideas were rare. Entropy played but a limited symbolic role in the degeneration ideology.

**The many faces of the ether**

According to most physicists in the second half of the nineteenth century, the world consisted not only of matter in motion but also, and no less importantly, of an all-pervading ethereal medium. The "luminiferous" ether was considered necessary to explain the propagation of light, and this was only one of the numerous purposes it served. In short, and in spite of a few dissenting voices, the ether was generally regarded indispensable in physics. The basic problem was not whether the ether existed or not, but the nature of the ether and its interaction with matter. Was the ether the fundamental substratum out of which matter was built? Or was matter a more fundamental ontological category of which the ether was just a special instance? The first view, where primacy was given to structures in the ether, came to be the one commonly held at the turn of the century and the years thereafter.

By that time the conception of the ether had changed, from a mechanical or hydrodynamic concept to one based on electrodynamics, but the change only made the ether even more important. In a textbook of 1902, Albert

Michelson, America's first Nobel laureate in physics, looked forward to the near future when all physical phenomena would be unified under a single theoretical framework. His optimism was rooted in "one of the grandest generalizations of modern science … that all phenomena of the physical universe are only different manifestations of the various modes of motion of one all-pervading substance – the ether" (Kragh 1999: 4).

From a technical point of view the ethereal world picture popular among Victorian physicists was not incompatible with the mechanical world picture, but it nonetheless differed from it by its emphasis on continuity rather than discreteness and the corresponding primacy given to ether over matter. According to the vortex atomic theory originally proposed by William Thomson in 1867, atoms were nothing but vortical structures in the continuous ether. In this sense the atoms were quasi-material rather than material bodies. As the ultimate and irreducible quality of nature, the ether could exist without matter, but matter could not exist without the ether. Or, as the distinguished theoretical physicist Joseph Larmor declared in his *Aether and Matter* published in 1900: "Matter may be and likely is a structure in the 'aether', but certainly aether is not a structure of matter" (p. vi; Noakes 2005: 422). Based on Thomson's idea, in the 1870s and 1880s the ambitious vortex theory evolved into a major research program. The theory was developed by several British physicists who examined it mathematically and applied it to a wide range of physical phenomena, including gravitation, the behavior of gases, optical spectra, and chemical combination (Kragh 2002). However, in the end this grand Victorian theory of everything proved unsuccessful.

By the early 1890s the vortex atomic theory had run out of steam and was abandoned by most researchers as a realistic theory of the constitution of matter. It was never unambiguously proved wrong by experiment, but after twenty years of work it degenerated into mathematics, failing to deliver what it promised of physical results. Physicists simply lost confidence in the theory. On the other hand, although the vortex atom was no longer considered a useful concept in explaining physical phenomena, heuristically and as a mental picture it lived on. Wrong as it was, to many British physicists it remained a methodological guiding principle, the ideal of what a future unified theory of matter and ether should look like. According to Michelson, writing in 1903, it "ought to be true even if it is not" (Kragh 2002: 80).

A contributing reason for the decline of the vortex theory was its foundation in the laws of mechanics, a feature that appeared unappealing in an environment increasingly hostile to the mechanical world view. Although the British vortex theorists presented their ether in a dematerialized form, they did not see it as entirely different from ordinary matter. It was mechanical in the sense that it possessed inertia as an irreducible property, and non-material only in the sense that it was continuous and not derivable from matter. From the perspective of many *fin-de-siècle* scientists, the emancipation from the thralldom of matter that vortex theory offered did not go far enough. At any rate, the demise of the vortex atom did in no way signal the demise of the ether. At the beginning of the new century the ether was much alive, believed to be as necessary as ever.

Although a physical concept and the basis of physical theory, the ether also served other purposes, in particular of an ideological and a spiritual kind (Noakes 2005). Larmor described it as "suprasensual." To some physicists, most notably Oliver Lodge, the ether became of deep spiritual significance, a psychic realm scarcely distinguishably from the mind. Not only was all nature emergent from the ether, he also came to see it as "the primary instrument of Mind, the vehicle of Soul, the habitation of Spirit … [and] the living garment of God" (Kragh 2002: 32). Lodge's extreme view was not shared by his fellow physicists, but it helped making the ether a popular concept among non-scientists and an ingredient of many *fin-de-siècle* speculations well beyond the limits of conventional science.

**Electrodynamics as world picture and world view**

As mentioned, the Zeitgeist of the *fin-de-siècle* physical sciences included a strong dose of anti-materialism, a desire to do away with brute matter and replace it with either energy or an ethereal medium. The demise of the mechanical ether models was followed by the emergence of a vigorous research program in which the ether was described by Maxwell's field theory of electromagnetism. Although Maxwell's theory dates from the 1860s, it was only in the last decade of the nineteenth century that physicists fully realized its amazing power. Among avant-garde physicists electromagnetism came to be seen as more fundamental than mechanics, a unifying principle of all science. It was assigned a role not unlike the one of energy in the energetics approach: in both cases, materialism was discarded and matter declared an epiphenomenon of a more basic entity, either energy or the electromagnetic field. The new electrodynamic approach proved more successful and progressive than the one of energetics, for other reasons because it fitted naturally within the ethereal world picture which it reinterpreted and breathed new life into. The replacement of the mechanical by the electromagnetic ether was arguably the most important change in fundamental physics in the years around 1900. While forgotten today, in the *fin-de-siècle* period it was regarded as a revolutionary advance in the understanding of nature (Jungnickel and McCormmach 1986: 227-244; Kragh 1999: 105-119).

Matter possesses mass, a fundamental quality that the vortex atom theory had been unable to explain in terms of the ether. Electrodynamics did better, for on the basis of Maxwell's theory a sphere of electricity could be assigned an "electromagnetic mass" with properties corresponding to those of the mass of ordinary matter. The question then arose of whether the mass of electrical particles could be completely accounted for in terms of electromagnetism, meaning that material or ponderable mass could be entirely disregarded. The question was considered in the 1890s by physicists such as Larmor and Lorentz who suggested the existence of "electrons" pictured as discrete structures in, or excitations of, the electromagnetic ether. These theoretical entities, localized ether in disguise, turned into real particles in 1897, when J. J. Thomson in Cambridge discovered negative electrons in a celebrated series of experiments with cathode rays. (To the surprise of the theorists, no positive electrons were detected.) By the closing years of the century, electrodynamics had given birth to electron physics, the conception of discrete subatomic particles wholly or partly of electromagnetic origin.

In a paper of 1900, significantly titled "On the Possibility of an Electromagnetic Foundation of Mechanics," the German physicist Wilhelm Wien outlined the basic features of a new research program the aim of which was to reduce all physical phenomena to electrodynamics. Five years later his compatriot Max Abraham referred to the program as the *electromagnetic world picture*, a name that indicates the theory's scope and ambitions. What, then, was the essence of this world picture or, in some versions, world view?

First of all, it was based on the belief that electrodynamics was more fundamental than mechanics, in the sense that the laws of mechanics could be fully understood electromagnetically. For example, Newton's laws of motion were merely special cases of the more profound and general laws governing the electromagnetic field. On the ontological level, it was claimed that there is nothing more to physical reality than what the science of electromagnetism tells us. All matter is made up of ethereal structures in the form of electrons, negative as well as positive. According to Augusto Righi, a prominent Italian physicist, electron theory was not so much an electromagnetic theory of matter as it was a replacement of matter by electromagnetism. "Matter is dead" was a common catchword in the early twentieth century. On the methodological level, the electromagnetic research program was markedly reductionistic, a theory of everything of a kind similar to the earlier vortex theory. The vision of the new and enthusiastic generation of electron physicists was, in a sense, that they were approaching the end of fundamental physics.

If mass is of electromagnetic origin it will increase with the speed or kinetic energy of the body in question, such as shown by Abraham, Lorentz and other electron theorists in the early twentieth century. It followed that the concepts of mass and energy could not be strictly separate, but that they must be connected by an equivalence relation of the same kind that Einstein famously proposed in 1905 (namely, $E = mc^2$). According to this point of view, matter was not really dead, it had merely metamorphosed into energy. Proposals of a mass-energy relationship predated Einstein's theory of relativity, and they added to the feeling that the entire foundation of physics had to be reconsidered. Young Einstein agreed, but for very different reasons. He saw no merit in the fashionable electromagnetic research program.

Although the ether was no less indispensable than in earlier theories, with the advent of electron physics its nature changed and became even more abstract and devoid of material attributes. Many physicists spoke of the ether as equivalent to the vacuum, or sometimes absolute space. According to a German electron theorist, August Föppl, the conception of space without ether was analogous to the contradictory conception of a forest without trees. However, this abstract and thoroughly dematerialized ether was more popular among German physicists than among their British colleagues. Just as FitzGerald had called for a physics "with a human interest," so J. J. Thomson, Lodge and Larmor tended to conceive the ether as a physical medium that played both a physical and metaphysical role. Lodge's ether was far from the abstract nothingness that some German physicists ascribed to the ethereal medium: in 1907 he calculated its energy density to no less than $10^{30}$ joule per cubic metre. The same year J. J. Thomson characterized the ether as an invisible universe that functioned as the workshop of the material universe. "The natural phenomena that we observe are pictures woven on the looms of this invisible universe," he said (Thomson 1908: 550).

By the early years of the twentieth century the electromagnetic view of the world had taken off and emerged as a highly attractive substitute for the outdated materialistic-mechanical view. Electron enthusiasts believed that physics was at a crossroad and that electrodynamics was on its way to establishing a new and possibly final paradigm of understanding nature. In 1904 the French physicist Paul Langevin gave voice to the revolutionary euphoria shared by at least a part of the physics community: "This [electromagnetic] idea has taken an immense development in the last few years, which causes it to break the framework of the old physics to pieces, and to overturn the established order of ideas and laws in order to branch out again in an organization which one foresees to be simple, harmonious, and fruitful" (Kragh 1999: 109).

The electromagnetic-ethereal world view, understood in a broad sense, was not limited to the small community of theoretical physicists. Consider the coming leader of world communism Vladimir Lenin, who in his *Materialism and*

*Empirio-Criticism* of 1908 discussed with sympathy and in some detail the new physical world view based on fields, ether and electrons (Kragh 1999: 110). His rhetoric was revolutionary, like Langevin's. Physics was in a crisis, but in a healthy crisis: "The entire mass of the electrons, or, at least, of negative electrons, proves to be totally and exclusively electrodynamic in origin. Mass disappears. The foundations of mechanics are undermined." Far from admitting that the new view of nature supported agnosticism and idealism, he saw it as solid support of materialism, although of the dialectical and not the mechanical version. "The electron," he declared, "is as *inexhaustible* as the atom, nature is infinite, but it infinitely *exists*." What he meant by this is not quite clear, but then perhaps it was not meant to be clear. It speaks to the pervasiveness of the electromagnetic world view, and the flexibility of dialectical materialism, that Lenin could persuade himself that the dematerialized, vacuum-like electromagnetic ether was in fact materialism on a higher level.

**Discoveries, expected and unexpected**

Physics at the turn of the century caused excitement not only because of the ambitious attempts to establish a new theoretical foundation, but also because of new discoveries that caught the physicists with surprise and contributed to the sense of crisis in parts of the physics community. While some of the discoveries could be interpreted within the framework of ether and electron physics, others stubbornly resisted explanation.

The experimental study of cathode rays from evacuated discharge tubes resulted in Thomson's electron, and it also led to Wilhelm Conrad Röntgen's X-rays announced in early 1896. Thomson's electron of 1897 was subatomic, with a mass about 1,000 times smaller than the hydrogen atom, and he originally thought of it as a *protyle*, the long-sought constituent of all matter. And this was not all, for he also believed that the electrons making up the cathode rays were produced by the dissociation of atoms. The cathode-rays tube worked as an alchemical laboratory! Atoms, he speculated, were not the immutable building blocks of matter, as traditionally believed, but could be transformed into the atoms of another element. On a cosmic time-scale, this was what had happened in nature. According to Thomson, an atom consisted of a large number of electrons in equilibrium positions, placed in and held together by a positive and weightless "fluid" of atomic dimension. This model, or something like it, was widely accepted in the early twentieth century.

While the electron had been anticipated theoretically, Röntgen's invisible yet penetrating rays were completely unexpected. They might be vibrations in the ether, or perhaps a particular species of electromagnetic waves, but their true nature was simply unknown. No less mysterious, and no less unexpected, was the phenomenon of radioactivity discovered by Henri Becquerel in Paris a few months later (Malley 2011). What was initially known as "uranium rays" grew out of studies of X-rays, and for a while physicists tended to conceive the new rays as merely a special kind of X-rays. It was only in 1898, with Marie and Pierre Curie's discovery of the highly active elements polonium and radium, that radioactivity became an exciting field of physical and chemical research, and also one that attracted great interest among physicians and the general public. Remarkably, radioactivity proved to be a spontaneous phenomenon that contrary to chemical reactions was independent of external conditions such as pressure, temperature and catalysts. Moreover, the property had been detected only for a few heavy elements. Might it still be a general property of matter? In the first decade of the twentieth century it was widely believed that all elements were radioactive, only most of them of such a low activity that it was difficult to measure.

Although radioactivity cannot be speeded up or slowed down, by 1902 it was established that the activity decreases over time at a characteristic rate or half-life given by the substance. According to the decay law found by Ernest Rutherford and Frederic Soddy, atoms of a particular substance decay randomly and in such a way that the probability is independent of the atom's age. The probability that a billion-year-old radium atom will decay within the next hour is exactly the same as that of a radium atom one second old.

According to Rutherford and Soddy, radioactivity was a material process, the spontaneous transmutation of one chemical element into another. The decay or disintegration hypothesis was controversial because it challenged the age-old dogma of the immutability of the elements, but within a year or two growing evidence forced most physicists and chemists to accept it. Combined with the belief that radioactivity is a common property of matter, it opened up for speculations about the ephemeral nature of all matter. Was matter gradually melting away? Into what? Soon after Rutherford and Soddy had shocked the world of science with their transmutation hypothesis, experiments proved that radioactive change is accompanied by large amounts of energy. From where did the energy come? What was the cause of radioactivity?

There was no shortage of answers, but they were all speculative, short-lived and unconvincing. In desperation a few physicists toyed with the idea that energy conservation might not be valid in radioactive processes. Others suggested that the missing energy came from unspecified ethereal rays or from the absorption of gravitation, and others again considered radioactivity as a kind of chemical reaction. However, the most accepted hypothesis was that the released energy was stored in the interior of the atom in the form of unstable configurations of electrons. The Thomson atom offered an explanation of sorts, if a qualitative one only. In 1905 an unknown Swiss physicist by the name Albert Einstein suggested that the source of energy was to be found in a loss of mass of the radioactive atoms. Few listened to him. In the decade following the discovery of radioactivity in 1896, the phenomenon remained deeply enigmatic and the source of countless speculations.

Not all scientists accepted the Rutherford-Soddy interpretation of radioactivity based on the transmutation of elements. As an instructive case, consider the eminent Russian chemist Dmitrii Mendeleev, the father of the periodic system (Gordin 2004: 207-238). A conservative and a realist in matters of chemistry and physics, he was disturbed by the new developments of *fin-de-siècle* physics such as the nonmaterial ether, the electron, the composite atom, and the evolution of elements based on a primordial form of matter. In works of 1903-1904 he warned that these developments might lead to all kinds of mystical and spiritual pseudoscience, leaving the material foundation of natural science behind. Mendeleev realized that the modern views of radioactivity and the composite atom made element transmutation a possibility, which to him was a sure sign of unscientific alchemy.

His worries were not unfounded. Alchemy, mixed up with elements of cabalism and spiritualism, experienced a revival at the turn of the century, often justifying its excessive claims by the results of the new physics (Morrisson 2007). In France, obscurantism flourished under the aegis of the Société Alchemique de la France and in England the spiritual and transcendental aspects of alchemy were cultivated within the Theosophical Society and later the Alchemical Society. Soddy and William Ramsay, two of Britain's foremost radiochemists and both of them Nobel laureates (of 1904 and 1921, respectively), described radioactivity in an alchemical terminology. If the transmutation theory of radium proved correct, said Ramsay in 1904, then "The philosopher's stone will have been discovered, and it is not beyond the bounds of possibility that it may lead to that other goal of the philosophers of the dark ages – the *elixir vitae*" (Morrisson 2007: 118). As Mendeleev saw it, there were good reasons to reject the modern transmutation theory of radioactivity.

**More mysterious rays**

Cathode rays, X-rays, and radioactivity were not the only kinds of rays that attracted attention in the years around 1900. In the wake of these discoveries followed several claims of new rays, most of which turned out to be spurious. The first decade of the new century witnessed the announcement of N-rays, rays of positive electricity, Moser rays, selenic rays, and magnetic rays, none of which exist. The N-rays that the French physicist René Blondlot claimed to have discovered in 1903 attracted much interest and were for a couple of years investigated by dozens of physicists, most of them French. The new rays (so named after Nancy, where Blondlot was professor) were emitted not only by discharge tubes but also from a variety of other sources such as the sun and gas burners. Sensationally, they seemed to be emitted also by the human nervous system, promising a connection between physics, physiology, and psychology. Although "seen" and examined by many physicists, by 1905 the consensus view emerged that N-rays do not exist. The effects on which the claim of the rays was based had no physical reality, but were of a psycho-sociological origin. Scientists saw them because they wanted to see them.

      In 1896 another Frenchman, the author, sociologist, and amateur physicist Gustave Le Bon, announced the discovery of what he called "black light," a new kind of invisible radiation that he believed was distinct from, but possibly related to, X-rays (Nye 1974). The discovery claim was welcomed by

French scientists, among them the great mathematician and physicist Henri Poincaré, many of whom supported Le Bon's general ideas of matter, radiation, and ether. Although black light turned out to be no more real than N-rays, for a while the discovery claim was taking seriously. In 1903 Le Bon was even nominated for a Nobel Prize in physics. Connected with the black light hypothesis, he developed a speculative, qualitative and time-typical theory of cosmic evolution and devolution which he presented in the best-selling *The Evolution of Matter* of 1906.

The chief message of Le Bon was that all matter is unstable and degenerating, constantly emitting rays in the form of X-rays, radioactivity and black light. Material qualities were held to be epiphenomena exhibited by matter in the process of transforming into the imponderable and shapeless ether from which it had once originated. The ether represented "the final nirvana to which all things return after a more or less ephemeral existence" (Le Bon 1907: 315). If all chemical elements emitted radioactive and similar ether-like rays, they would slowly melt away, thereby proving that matter could not be explained in materialistic terms. Energy and matter were two sides of the same reality, he declared, different stages in a grand evolutionary process that in the far future would lead to a kind of heat death, a pure ethereal state. However, contrary to the thermodynamic heat death, Le Bon's ethereal nirvana might not be the final end state of the universe: he suggested that it would perhaps be followed by a new cosmic birth and evolution, and that the cyclic process might go on eternally.

Although clearly speculative, many scientists found Le Bon's ideas attractive or came independently to somewhat similar cosmic scenarios. The views of Lodge in England were in many ways congruent with those of Le Bon. In both cases, the views appealed to the anti-materialist, evolutionary, and holistic sentiments that were popular in the *fin-de-siècle* period. It was part of the Zeitgeist, both in France and elsewhere, that many scientists were willing to challenge established knowledge, including doctrines such as the permanence of the chemical elements and the law of energy conservation. The very qualities of permanence and conservation were considered suspicious within an intellectual climate emphasizing transformation, evolution, and becoming. The ideas of the American polymath Charles Sanders Peirce, trained in physics and astronomy but today better known as a philosopher, may serve to illustrate the *fin-de-siècle* themes of contingency and indeterminism. Peirce's system of science, which he published in a series of papers 1891-1893, was thoroughly indeterministic and probabilistic. It assumed that the laws of nature were only statistically valid and left open the possibility that they might even be evolutionary, that is, vary in time (Peirce 1966).

Part of the intellectual climate at the time tended toward anti-science, or at least anti-scientism, in so far that it contrasted the scientific world view with an idealistic understanding of the world that included irrational, emotional, and spiritual perspectives. Le Bon's quasi-scientific speculations had considerable appeal among those who were dissatisfied with positivistic ideals and longed for an undogmatic, more youthful science that would better satisfy what they associated with the human spirit. His ideas struck a chord in a period that has been described as a "revolt against positivism" and in which science was charged with being morally bankrupt (MacLeod 1982). Although far from anti-science, Le Bon joined the trend and flirted with its values. According to a French newspaper of 1903, "Poincaré and Le Bon fearlessly undermine the old scientific dogmas [and] do not fear saying that these cannot fulfill and satisfy the human spirit. We recognize along with these teachers … the bankruptcy of science" (Nye 1974: 185). Whereas Poincaré and many other French scientists listened with sympathy to Le Bon's arguments, he was not taken seriously among German physicists. The reason may have been that his ether was outdated, as he failed to incorporate the electromagnetic ether in his speculations.

In *The Value of Science*, a collection of articles from 1905, Poincaré argued that physics was in a state of crisis, not only because of the uncertainty with respect to mass and energy conservation, the problems with the second law of thermodynamics, and the new electron theory, but also because of the mysterious radium – "that grand revolutionist of the present time," as he called it. Lenin, in his *Materialism and Empirio-Criticism*, approvingly quoted Poincaré, and he was not the only one to consider radioactivity revolutionary in more than the scientific sense. The disintegration hypothesis, and radioactivity generally, was sometimes seen as subversive not only to the established scientific world view but also to the political order. According to the Spanish physicist and intellectual José Echegaray, radium appeared as "a revolutionary metal, like an anarchist that comes to disturb the established order and to destroy all or most of the laws of the classical science" (Herran 2008: 180).

**Spritualism and hyperspace**

As mentioned, there was in some parts of the *fin-de-siècle* scientific community a tendency to extend results of the new physical world picture to areas of a non-scientific nature, such as alchemy, occultism, spiritualism, and related paranormal beliefs. The great decade of spiritualism was the 1870s, after which it waned somewhat, but still at the turn of the century it continued to attract the interest of many physical scientists, including some of the highest distinction. In England, the Society for Psychical Research had been founded in 1882 and among its members were luminaries such as Lord Rayleigh, J. J. Thomson,

William Crookes, and Lodge. Membership of the society indicated an interest in the psychic or spiritual realm, but not necessarily a belief in the reality of psychic phenomena (Oppenheim 1985). Among the skeptics were Rayleigh and Thomson, whose attitude derived as much from their Christian belief as from scientific reasons. In spite of different attitudes, the scientific members of the society agreed that spirit should not automatically be banished from the world revealed by modern science.

Although the standard view then as now was to consider spiritualism antithetical to science, many believers argued that séances with psychic media provided *scientific* evidence for the survival of the spirit after bodily death. Neither did the majority of scientists perceive science as inherently inimical to spiritualist belief. They thought that the reality of psychic phenomena could be examined by the ordinary critical methods of science. According to Lodge, human immortality was not a working hypothesis but a scientific fact. But there were also those who reached the conclusion that science and spiritualism, although not in conflict, could not be reconciled either: the study of spiritualism was legitimate in its own right, as based on religious and psychological considerations, but there was no such thing as spiritualist science.

With few exceptions, physical scientists in the Society for Psychical Research avoided explaining spiritual phenomena by invoking the ether, electromagnetic forces, radioactive transmutation, or other parts of modern physics. Yet some did, such as the chemist and physicist Crookes who had become convinced of the reality of telepathy. In an address of 1897 he vaguely suggested that telepathic powers might be explained by X-rays or something like them acting on nervous centers in the brain. From a different perspective, in 1902 a Spanish socialist magazine explained to its readers that radioactivity was likely to do away with supernatural causes of telepathy and like paranormal phenomena, which instead could be explained on a sound physical basis (Herran 2008).

Contrary to some of his colleagues in England, the Leipzig professor in physics and astronomy Carl Friedrich Zöllner was convinced that spiritualism belonged to the realm of science. Inspired by Crookes, in the years around 1880 he investigated in great detail the spiritual world in séances with participation of leading German scientists and philosophers (Treitel 2004). A believer in the reality of spiritualist manifestations, Zöllner published in 1879 his *Transcendental Physics*, a book that appeared in several editions in both German and English. To his mind, the project of a transcendental physics including both material and spiritual phenomena was but a natural extension of the astrophysical project of accommodating terrestrial and celestial phenomena within the same theoretical framework.

The distinguishing feature in Zöllner's transcendental physics was the crucial role played by a hypothetical fourth dimension of space as the site of paranormal phenomena. He was convinced that the reality of four-dimensional space could be established experimentally, indeed that there was already incontrovertible scientific evidence for it. This extended space was transcendental but nonetheless subject to physical analysis, and it was destined to revolutionize physics. Zöllner and his followers argued that there were natural phenomena that defied causal explanation in three-dimensional space, and that they could be understood only in terms of forces from a higher dimension.

Ideas of a four-dimensional "hyperspace" were common at the end of the nineteenth century, if rarely entertained by leading scientists and in most cases without the direct association to spiritualism that Zöllner advocated. The English-American mathematician and author Charles Howard Hinton wrote a series of articles and books, including *The Fourth Dimension* of 1904, in which he claimed that the extra space dimension might explain physical phenomena such as the nature of ether and electricity. He suggested that the three-dimensional mechanical ether formed a smooth sheet in the fourth dimension, the thickness of the sheet being approximately that of "the ultimate particles of matter." Among the few scientists of distinction who entertained similar ideas was the American astronomer Simon Newcomb, who in 1896 uncommittedly speculated that "Perhaps the phenomena of radiation and electricity may yet be explained by vibration in a fourth dimension" (Beichler 1988: 212).

Although hyperspace models of the ether and similar speculations of a fourth dimension were well known in the *fin-de-siècle* period, they were peripheral to mainstream developments in physics, such as the turn from mechanics to electrodynamics. The fourth dimension caught the public imagination, was eagerly adopted by occultists and idealist philosophers, and became an important utopian theme in literature and art in the early twentieth century (Henderson 2009). Most physicists considered it a harmless speculation of no scientific use.

**Towards modern physics**

The spirit of physics in the *fin-de-siècle* period was both conservative and revolutionary. At the same time as the majority of physicists worked within the framework of Newtonian mechanics, there was a growing dissatisfaction with this framework, in part for scientific reasons and in part for reasons that may best be described as ideological. The dissatisfaction resulted in what has been called a "neoromantic" trend (Brush 1978), attempts to establish the physical sciences on a new foundation that did not share the elements of materialism with which the older paradigm was often associated. As a result, classical physics was partially overthrown, but without a new "modern" physics taking its place. The great unifying concepts of the new physics were energy and ether, the latter described by the equations of electromagnetism rather than mechanics. Whether in the form of energetics or the electrodynamic world picture, the great expectations of the neoromantic physicists were not fulfilled. The revolution failed or rather, it was overtaken by another and at the time hardly visible revolution that was antithetical to both the classical world view and the one envisioned by the *fin-de-siècle* physicists.

The two pillars of this revolution – what today is recognized at the beginning of modern physics – were Planck's quantum theory of 1900 and Einstein's theory of special relativity dating from 1905. Both theories stood apart from the fashionable theories of the time based on ether, energy, and electrodynamics. Einstein's theory operated with four dimensions, but these were quite different from the earlier hyperspace speculations: in relativity theory it is spacetime and not space that has four dimensions. With the recognition of the theories of quanta and relativity the ethereal world view so characteristic of *fin-de-siècle* physics became obsolete. Einstein summarily dismissed the ether as an unnecessary construct, and the new theory of quanta proved incompatible with the electrodynamic ether embraced by avant-garde physicists. As seen in retrospect, the enduring and really important contributions of the *fin-de-siècle* period to modern physics were not the ambitious attempts to create a new unified foundation of physics. They are rather to be found in the period's experimental discoveries of, for example, X-rays, the electron, and radioactivity. These discoveries were initially interpreted within the framework of *fin-de-siècle* physics, but, with the possible exception of the electron, they were not products of it.